\documentclass[aps,prl,twocolumn,groupedaddress,floatfix,preprintnumbers]{revtex4}
\usepackage[centertags]{amsmath}
\usepackage{amssymb}
\usepackage{amsmath,multirow,array}
\usepackage{graphicx}
\usepackage[usenames]{xcolor}

\usepackage{units}

\begin{document}

\title{Universality of second order transport in Gauss-Bonnet gravity}
\author{Evgeny Shaverin}
\author{Amos Yarom}
\affiliation{Department of Physics, Technion, Haifa 32000, Israel}

\date{\today}

\begin{abstract}
We compute all the second order transport coefficients of a hydrodynamic theory with a gravity dual which includes a Gauss-Bonnet term. We find that a particular linear combination of the second order transport coefficients, which was found to vanish in generic two derivative gravity theories with matter, remains zero even in the presence of the Gauss-Bonnet term. We contrast this behavior with the shear viscosity to entropy density ratio.

\end{abstract}

\pacs{}

\maketitle

\textit{Introduction.}---The AdS/CFT correspondences relates the strongly coupled phase of $3+1$ dimensional conformal gauge theories (CFT's) to theories of gravity compactified to $4+1$ dimensions \cite{Maldacena:1997re,Witten:1998qj,Gubser:1998bc}. In the strongly coupled planar limit, the rank of the gauge-group $N$ and its 't~Hooft coupling $\lambda$ are related to Newtons constant $G_{N}$ and the string scale $\alpha'$ via
\begin{equation}
\label{E:dictionary1}
	\lambda \propto \frac{L^4}{\alpha^{\prime\,2}}
	\qquad
	N^2 \propto \frac{L^3}{G_N}
\end{equation} 
where the exact coefficients in the relations \eqref{E:dictionary1} depend on the details of the theory \cite{Gubser:1998vd}.
In this work we will be interested in the hydrodynamic phase of non-charged strongly coupled gauge theories whose dual is given by a neutral, asymptotically AdS black hole. In particular, we will be interested in corrections to the second order transport coefficients of the dual gauge theory which are induced by Gauss-Bonnet corrections to the gravitational action.

Adding a  Gauss-Bonnet term to the gravitational action may be thought of as an effective contribution to the action which arises from a variety of possible stringy corrections. For instance, it could capture some of the effects of closed string loop corrections  \cite{Anselmi:1998zb}
or it could be induced by orientifold planes or D branes which in certain instances would correspond to changing the gauge group from the canonical $SU(N)$ to $SO(N)$ or $USp(2N)$ \cite{Fayyazuddin:1998fb,Aharony:1998xz,Aharony:1999rz}.
Regardless of its origin, the appearance of a Gauss-Bonnet term in the gravitational induces a shift in the central charges $a$ and $c$ of the CFT as we now explain.

The trace anomaly of a CFT can be parameterized by central charges $a$ and $c$ such that
\begin{align}
\label{E:canda}
	T^{\mu}_{\mu} =& \frac{c}{16 \pi^2} \left(R_{\mu\nu\rho\sigma}R^{\mu\nu\rho\sigma} - 2 R_{\mu\nu}R^{\mu\nu} + \frac{1}{3}R^2\right) \nonumber \\
		&-\frac{a}{16\pi^2} \left(R_{\mu\nu\rho\sigma}R^{\mu\nu\rho\sigma} - 4 R_{\mu\nu}R^{\mu\nu} + R^2\right)
\end{align}
with $T^{\mu}_{\mu}$ the trace of the stress tensor and $R_{\mu\nu\rho\sigma}$, $R_{\mu\nu}$ and $R$ the Riemann tensor, Ricci tensor and Ricci scalar respectively. In the absence of higher derivative corrections to the gravitational action one has $a=c$ \cite{Henningson:1998gx,Gubser:1998vd}. The introduction of a Gauss-Bonnet term to the action implies that $a-c \neq 0$ \cite{Blau:1999vz,Buchel:2008vz}. In what follows, we will assume that there exists a regime of parameters of the theory in which one can consistently neglect six and higher derivative corrections to the gravitational action. (See for instance \cite{Kats:2007mq,Buchel:2008vz} for explicit examples where such conditions may be satisfied.) In this strongly coupled regime, we will study corrections to second order transport coefficients of the hydrodynamic phase of the dual gauge theory.

Hydrodynamics can be thought of as an effective theory, which, in the absence of conserved charges, is characterized by a velocity field $u^{\mu}$ normalized such that $u^{\mu}u_{\mu} = -1$ and a temperature $T$. Expanding the energy momentum tensor to second order in gradients of the hydrodynamic variables and imposing Weyl covariance and tracelessness of the energy momentum tensor, one finds
\begin{equation}
T_{\mu\nu}=P(4u_{\mu}u_{\nu}+\eta_{\mu\nu})-\eta\sigma_{\mu\nu}+\sum_{i=0}^{3}\lambda_{i}\Sigma_{\mu\nu}^{\left(i\right)}
\end{equation}
where
\begin{equation}
\sigma_{\mu\nu}=2\partial_{\langle\mu}u_{\nu\rangle}\,,\quad\omega_{\mu\nu}=\frac{1}{2}P_{\mu}{}^{\alpha}P_{\nu}{}^{\beta}\left(\partial_{\alpha}u_{\beta}-\partial_{\beta}u_{\alpha}\right)
\end{equation}
and
\begin{align}
\begin{split}
	\Sigma^{(0)}_{\mu\nu} &= {}_{\langle}u^{\alpha}\partial_{\alpha}\sigma_{\mu\nu\rangle} + \frac{1}{3} \sigma_{\mu\nu}\partial_{\alpha}u^{\alpha} \\
	\Sigma^{(1)}_{\mu\nu} &= \sigma_{\langle \mu \alpha}\sigma^{\alpha}{}_{\nu\rangle} \,,
	\quad
	\Sigma^{(2)}_{\mu\nu} = \sigma_{\langle \mu\alpha} \omega^{\alpha}{}_{\nu\rangle} \,,
	\quad
	\Sigma^{(3)}_{\mu\nu} = \omega_{\langle \mu\alpha} \omega^{\alpha}{}_{\nu\rangle}
\end{split}
\end{align}
and triangular brackets denote a traceless transverse projection
\begin{equation}
	A_{\langle \mu \nu \rangle}= \frac{1}{2} P_{\mu}{}^{\alpha}P_{\nu}{}^{\beta} \left(A_{\alpha\beta}+A_{\beta\alpha}\right) - \frac{1}{3}P_{\mu\nu}P^{\alpha\beta}A_{\alpha\beta}
\end{equation}
with
\begin{equation}
	P^{\mu\nu} = \eta^{\mu\nu} + u^{\mu}u^{\nu}\,.
\end{equation}
See \cite{Baier:2007ix,Bhattacharyya:2008jc} for more details regarding the derivative expansion.

In \cite{Buchel:2003tz}, following \cite{Policastro:2001yc,Policastro:2002se}, it was shown that in theories which preserve rotational invariance, the shear viscosity to entropy density ratio $\eta/s$ satisfies 
\begin{equation}
\label{E:etas}
	\frac{\eta}{s} = \frac{1}{4\pi}
\end{equation}
in the supergravity limit, regardless of the matter content of the theory. (When rotational symmetry is broken then \eqref{E:etas} no longer holds as has been shown in \cite{Erdmenger:2011tj,Rebhan:2011vd}.) Taking the result \eqref{E:etas} and applying it unwaveringly to the quark gluon plasma which is presumably created in the process of a collision of two heavy ions gives results which are in qualitative agreement with experiment \cite{CasalderreySolana:2011us}. 

When computing second order transport coefficients of a conformal theory using the gauge-gravity duality, one obtains a similar relation \cite{Erdmenger:2008rm,Haack:2008xx}
\begin{equation}
\label{E:linear}
	- 2 \lambda_0 + 4 \lambda_1  - \lambda_2 = 0\,.
\end{equation}
The role of second order transport coefficients in simulations of heavy ion collisions has been discussed in, for example, \cite{CasalderreySolana:2011us}.

Once Gauss-Bonnet corrections to Einstein-gravity are taken into account, relation \eqref{E:etas} breaks down and one finds, instead,
\begin{equation}
\label{E:etasd}
	\frac{\eta}{s} = \frac{1}{4\pi}\left(1- \delta + \mathcal{O}(\delta^2) \right)
\end{equation}
where $\delta =  \frac{c-a}{c}$ and $c$ and $a$ correspond to the central charges of the CFT as defined in \eqref{E:canda} \cite{Brigante:2007nu,Brigante:2008gz,Kats:2007mq,Buchel:2008vz} (see also \cite{Buchel:2008ac,Dutta:2008gf,Brustein:2008cg,Cai:2009zv,Cremonini:2009sy,Myers:2009ij,Banerjee:2009wg,Banerjee:2009fm,Cremonini:2012ny,Brustein:2012pq} for related work). 
In what follows, we show that the relation \eqref{E:linear} does not receive corrections in the presence of Gauss-Bonnet terms and remains valid at least to order $\mathcal{O}(\delta^2)$.

As pointed out in \cite{Haack:2008xx} the relations \eqref{E:etas} and \eqref{E:linear} while similar in spirit do differ in a consequential way. While \eqref{E:etas} relates transport properties $\eta$ to equilibrium properties $s$, equation \eqref{E:linear} is a relation among transport coefficients. 

The deviation \eqref{E:etasd} of the ratio of the shear viscosity to entropy density from the value given in \eqref{E:etas} is, perhaps, expected since \eqref{E:etas} does not hold in the non planar weakly coupled theory. Turning our attention to \eqref{E:linear}, the value of the $\lambda_i$'s at weak coupling were computed for various theories in \cite{York:2008rr} using kinetic theory, with results which deviate from \eqref{E:linear}. Since $a$ and $c$ do not get corrected by marginal parameters \cite{Anselmi:1998zb}, we expect that the difference $c-a$ which controls the coefficient of the Gauss Bonnet term is not associated with stringy, $\alpha'$, corrections.  It would be interesting to study the effect of six and higher order corrections to the gravitational action (which should correspond to subleading corrections to the t' Hooft coupling) on equation \eqref{E:linear} (see \cite{Saremi:2011nh} for initial progress in this direction).
\\ \phantom{} \\
\textit{Computation}---The starting point for our computation is the action
\begin{equation}
	S = -\frac{1}{16 \pi G_5} \int \sqrt{-g} \left( R + \frac{12}{L^2} - \theta L_{GB} \right) d^5x + S_b
\end{equation}
where
\begin{equation}
	 L_{GB} = R_{mnpq}R^{mnpq} - 4 R_{mn}R^{mn} + R^2
\end{equation}
and $S_b$ are appropriate counter terms which make the variational principle well defined and the Brown-York stress tensor finite \cite{deHaro:2000xn,Myers:1987yn}. The Roman indices $m,\,n=0,\ldots,4$ refer to bulk quantities while Greek indices $\mu,\,\nu = 0,\ldots,3$ refer to boundary quantities. The parameter $\theta$ which controls the strength of the Gauss-Bonnet term is related to the field theory quantity $\delta= (c-a)/a$ via $8 \theta = \delta$ \cite{Blau:1999vz,Buchel:2008vz}.

In a coordinate system where the asymptotically AdS geometry is given by 
\begin{equation}
	\lim_{r \to \infty} ds^2 = r^2 \eta_{\mu\nu} dx^{\mu}dx^{nu}\,,
\end{equation}
The prescription for computing the boundary theory stress tensor $T_{\mu\nu}$ to linear order in $\theta$ is given by \cite{Brihaye:2008kh,Dutta:2008gf}
\begin{flalign}
\label{E:Tmnprescription}
	& T_{\mu\nu}   = 
		 \lim_{r\to\infty}\frac{r^{2}}{L^{2}8\pi G_5} \times \\
	&\left(\mathcal{K}_{\mu\nu}-\mathcal{K}\gamma_{\mu\nu}
		+2\theta\left(3\mathcal{J}_{\mu\nu}-\mathcal{J}\gamma_{\mu\nu}\right)-\frac{3}{L}\gamma_{\mu\nu}+\frac{\theta}{L^{3}}\gamma_{\mu\nu}\right) \nonumber
\end{flalign}
where
\begin{equation}
	\gamma_{mn} = g_{mn} - N_m N_n
\end{equation}
is the boundary metric with $N_n = \delta_n^4 / \sqrt{g^{44}}$ a unit outward vector to the boundary and 
\begin{equation}
	\mathcal{K}_{mn} = -\frac{1}{2} \left(\nabla_{m} N_n + \nabla_n N_m\right)
\end{equation}
is the extrinsic curvature on the boundary. The tensor $\mathcal{J}_{mn}$ is given by
\begin{multline}
\mathcal{J}_{mn} = \frac{1}{3} \Big(2 \mathcal{K} \mathcal{K}_{mp}\mathcal{K}^p_{\phantom{p}n} + \mathcal{K}_{ps}\mathcal{K}^{ps} \mathcal{K}_{mn} - \\
	2 \mathcal{K}_{mp}\mathcal{K}^{ps}\mathcal{K}_{sn} - \mathcal{K}^2 \mathcal{K}_{mn}\Big)
\end{multline}
and $\mathcal{K}$ and $\mathcal{J}$ are the trace of $\mathcal{K}_{mn}$ and $\mathcal{J}_{mn}$ respectively.

The resulting equations of motion are given by:
\begin{flalign}
	& R_{mn}-\frac{1}{2}Rg_{mn}-\frac{6}{L^{2}}g_{mn}-\frac{\theta}{2}g_{mn}L_{GB}  \\
	& +2\theta\left(R_{mpql}R_{n}^{\phantom{n}pql}-2R^{pq}R_{mpnq}-2R_{m}^{\phantom{m}q}R_{qn}+RR_{mn}\right)=0 \,.
	\nonumber
\end{flalign}
If we use an ansatz of the form
\begin{equation} \label{eq:ansatz0}
	ds^2 = -r^2 f dt^2 + r^2 (dx^i)^2 + 2 S dt dr
\end{equation}
and set $L=1$, then we find that
\begin{equation}
	f = 1 - \frac{1}{b^4 r^4} + \frac{2 \theta}{b^8r^8}
	\qquad
	S = 1 - \theta
\end{equation}
solves the equations of motion to linear order in $\theta$. Here $b$ is a conveniently chosen integration constant. The resulting energy momentum tensor which follows from the prescription \eqref{E:Tmnprescription} is given by
\begin{equation}
	T_{\mu\nu} = \frac{T^4 \pi^4}{16 \pi G_{5}} (1+3\theta) \, \hbox{diagonal}(3,\,1,\,1,\,1) 
\end{equation}
where $T$ is the Hawking temperature, related to $b$ through
\begin{equation}
	T = \frac{2-3 \theta}{2 b \pi} \,,
\end{equation}
and $G_5$ is related to the rank of the gauge group through a relation of the form \eqref{E:dictionary1}.

To compute the transport coefficients $\eta$, and $\lambda_i$ we follow the prescription of \cite{Bhattacharyya:2008jc}. We boost the solution \eqref{eq:ansatz0} so that it takes the form
\begin{equation}
	ds^2 = -r^2 f u_{\mu}u_{\nu} dx^{\mu}dx^{\nu} + r^2 P_{\mu\nu} dx^{\mu}dx^{\nu} - 2 S u_{\mu} dx^{\mu}dr
\end{equation}
where
\begin{equation}
	u^{\mu}=\frac{1}{\sqrt{1-\beta^{2}}}\left(1,\vec{\beta}\right)\,
	\quad
	P_{\mu\nu} = \eta_{\mu\nu} + u_{\mu}u_{\nu}\,.
\end{equation}
We now promote the integration constants $\beta_i$ and $b$ to become space-time dependent fields $\beta_i(x^{\alpha})$ and $b(x^{\alpha})$ and correct the metric order by order in derivatives of $u^{\mu}$ and $b$. It is most efficient to decompose the metric into scalar, vector and tensor modes of the $SO(3)\subset SO(3,1)$ symmetry under which $u^{\mu}$ is (locally) invariant, i.e., we write
\begin{flalign}
\label{E:lineelement}
	ds^{2} & =r^{2}ku_{\mu}u_{\nu}dx^{\mu}dx^{\nu}+r^{2}P_{\mu\nu}dx^{\mu}dx^{\nu}-2Su_{\mu}dx^{\mu}dr \nonumber \\
		& \phantom{=}+r^{2}\left(u_{\mu}V_{\nu}+u_{\nu}V_{\mu}\right)dx^{\mu}dx^{\nu}+r^{2}\Pi_{\mu\nu}dx^{\mu}dx^{\nu}
\end{flalign}
and expand $k$, $V$, $S$ and $\Pi$ in gradients of $u^{\mu}$ and $b$. It is convenient to denote the $n$'th order correction to $k$, $V$, $S$ and $\Pi$ with a superscript $(n)$. Thus, for example
\begin{equation}
	\Pi^{(0)}_{\mu\nu} = V^{(0)}_{\mu} = 0
	\qquad
	S^{(0)} = 1-\theta
	\qquad
	k^{(0)} = f\,.
\end{equation}

In what follows we will present our computation using the dimensionless parameter $\rho = r b$ in favor of $r$. For $n\geq 1$, and to first order in $\theta$, we find that the equations of motion for $S^{(n)}(\rho)$, $k^{(n)}(\rho)$, $V^{(n)}_{\mu}(\rho)$ and $\Pi^{(n)}_{\mu\nu}(\rho)$ take the form 
\begin{flalign}
\begin{split}
\label{E:EOMs}
	\left(S^{(n)}\right)^{\prime} & =\mathbf{S}^{(n)}\\
	\left(\left(\rho^{4}+4\theta\right)k^{(n)}\right)^{\prime}\qquad\qquad\qquad\qquad\qquad\\
		+2\left(\left(\rho^{4}-1+\theta\left(3-\frac{2-\rho^{8}}{\rho^{4}}\right)\right)S^{(n)}\right)^{\prime} 
		& =\mathbf{k}^{(n)}\\
	\left(\rho\left(\rho^{4}+4\theta\right)\, V_{\mu}^{(n)\,\prime}\right)^{\prime} & =\mathbf{V}_{\mu}^{(n)}\\
	\left(\left(\rho(\rho^{4}-1)+2\theta\left(\frac{3-2\rho^{4}}{\rho^{3}}\right)\right)\,
		\Pi_{\mu\nu}^{(n)\,\prime}\right)^{\prime} & =\mathbf{P}_{\mu\nu}^{(n)}
\end{split}
\end{flalign}
where ${}^\prime$ denotes a derivative with respect to $\rho$ and $\mathbf{S}^{(n)}(\rho)$, $\mathbf{P}^{(n)}_{\mu\nu}(\rho)$, $\mathbf{k}^{(n)}(\rho)$ and $\mathbf{V}^{(n)}_{\mu}(\rho)$ are source terms which have to be determined perturbatively. We emphasize that \eqref{E:EOMs} are correct up to $ \mathcal{O}\left(\theta^{2}\right) $. 

The solution to \eqref{E:EOMs} is dual to a hydrodynamic state in the boundary theory. Consequentially, the boundary conditions we impose when solving \eqref{E:EOMs} are that the fields are finite at the horizon ($\rho=1$), that the boundary theory metric is the Minkowski metric (implying that the fields $S^{(n)}$, $k^{(n)}$, $V_{\mu}^{(n)}$ and $\Pi_{\mu\nu}^{(n)}$ fall off fast enough near the boundary) and, in addition, we require that the boundary theory stress tensor defined in \eqref{E:Tmnprescription} is in the Landau frame (implying that the $\mathcal{O}(r^{-4})$ component of $k^{(n)}$ and $V_{\mu}^{(n)}$ vanish at the asymptotic AdS boundary). We refer the reader to \cite{Bhattacharyya:2008jc,Haack:2008cp} for a more detailed description of these boundary conditions. 

After solving the equations of motion we can compute the hydrodynamic stress tensor by inserting \eqref{E:lineelement} into \eqref{E:Tmnprescription} and expanding to $\mathcal{O}(\theta^2)$ and $\mathcal{O}(\partial^3)$. After some algebra, we find that
\begin{equation}
\label{E:Tmnprescription2}
	T_{\mu\nu}= \frac{T^4 \pi^4}{16 \pi G_{5}} (1+3\theta)(4u_{\mu}u_{\nu}+\eta_{\mu\nu})  +\frac{\left(1-3\theta\right)}{4\pi G_{5}}\pi_{\mu\nu}\,,
\end{equation}
where $ \pi_{\mu\nu}$ is the coefficient of the $\mathcal{O}(r^{-4})$ term of $ \Pi_{\mu\nu} $ at large $r$.  (An explicit computation of the $\theta=0$ limit of \eqref{E:Tmnprescription2} can be found in \cite{Haack:2008cp}.)

The solution to the equations of motion to first order in gradients was computed in \cite{Dutta:2008gf}. For completeness we reproduce it here. The sources for the equations of motion take the form
\begin{align}
\begin{split}
	\mathbf{S}^{(1)} & =0\\
	\mathbf{k}^{(1)} & =b\left(2\rho^{2}-\theta\frac{6\rho^{4}+8}{3\rho^{2}}\right)\partial_{\alpha}u^{\alpha}\\
	\mathbf{V}_{\mu}^{(1)} & =b\left(3\rho^{2}-
		\theta\frac{3\rho^{4}+4}{\rho^{2}}\right)u^{\alpha}\partial_{\alpha}u_{\mu}\\
	\mathbf{P}_{\mu\nu}^{(1)} & =b\left(-3\rho^{2}+\theta\frac{3\rho^{4}-4}{\rho^{2}}\right)\sigma_{\mu\nu}
\end{split}
\end{align}
and the solution is given by
\begin{align}
\begin{split}
	S^{(1)}(r)  &= 0 \\
	k^{(1)}(r)  &= \frac{2}{3r}(1-\theta) \, \partial_\alpha u^\alpha \\
	V_{\mu}^{(1)}(r)  &=  - \frac{1}{r}(1-\theta) \, u^\alpha \partial_\alpha u_\mu \\
	\Pi_{\mu\nu}^{(1)}(r) &= \left[ \frac{1}{2} (2 + 3 \theta) \, b F_1(br) + \theta \, b F_2(br) \right] \sigma_{\mu\nu}
\end{split}
\end{align}
where
\begin{align*}
F_1(\rho) &= -\ln(\rho) + \frac{1}{2}\ln(\rho + 1) + \frac{1}{4}\ln(\rho^2 + 1)  \\
	& \quad - \frac{1}{2} \arctan(\rho) + \frac{\pi}{4} \\
F_2(\rho) &= \frac{1}{4} \left( \frac{6}{\rho^4} - \frac{8}{\rho} - \frac{1}{1+\rho} - \frac{1+\rho}{1+\rho^2} \right)
\end{align*}

Using \eqref{E:Tmnprescription2} , the resulting hydrodynamical energy momentum tensor is given by
\begin{equation}
	T_{\mu\nu} = \frac{T^4 \pi^4}{16 \pi G_{5}} \left( 1+3 \theta \right) \left(4 u_{\mu} u_{\nu} + \eta_{\mu\nu} \right)
		- \frac{T^3 \pi^3}{16 \pi G_{5}} \left( 1-5 \theta \right) \sigma_{\mu\nu} \,
\end{equation}
from which we can read the shear viscosity
\begin{equation}
	\eta = \frac{T^3 \pi^3}{16 \pi G_{5}} \left( 1-5 \theta \right) \,.
\end{equation}
We can evaluate the entropy density of the system using $s = dP/dT$ to obtain
\begin{equation}
	\frac{\eta}{s} = \frac{1}{4 \pi} (1 - 8 \theta) = \frac{1}{4\pi}(1-\delta)
\end{equation}
which reproduces the results in \cite{Brigante:2007nu,Dutta:2008gf}.

Since we are working in the Landau frame then, according to \eqref{E:Tmnprescription2}, in order to compute the second order transport coefficients we need only solve the tensor equations for $\Pi_{\mu\nu}^{(2)}$ in \eqref{E:EOMs}. An explicit computation gives us
\begin{equation}
	\mathbf{P}^{(2)}_{\mu\nu} = \mathbf{P}_0^{(2)} \Sigma^{(0)}_{\mu\nu} + \mathbf{P}_1^{(2)} \Sigma^{(1)}_{\mu\nu} + \mathbf{P}_2^{(2)} \Sigma^{(2)}_{\mu\nu} + \mathbf{P}_3^{(2)} \Sigma^{(3)}_{\mu\nu}
\end{equation}
where
\begin{widetext}
\begin{flalign*}
	b^{-2}\mathbf{P}_{0}^{(2)}(\rho) &= \rho-2\rho^{3/2}\left(\rho^{3/2}F_{1}\right)^{\prime}
		+\theta\Biggl[-2\rho+\frac{12}{\rho^{3}}-2\rho^{3/2}\left(\rho^{3/2}\left(2F_{1}+F_{2}\right)\right)^{\prime}
		+\left(\frac{8+3\rho^{4}}{\rho}\right)^{\nicefrac{1}{2}}\left(\left(\frac{8+3\rho^{4}}{\rho}\right)^{\nicefrac{1}{2}}F_{1}\right)^{\prime}\Biggr]\\
	b^{-2}\mathbf{P}_{1}^{(2)}(\rho) &= \rho-3\rho^{2}F_{1}+\rho\left(\rho^{4}-1\right)F_{1}^{\prime}{}^{2}
			+\theta\Biggl[-6\rho+\frac{12}{\rho^{3}}-\left(\frac{8+3\rho^{4}}{2\rho^{2}}\right)F_{1}
			+2\rho\left(\rho^{4}-1\right)F_{1}^{\prime}\,\left(F_{1}^{\prime}+F_{2}^{\prime}\right)\\
		& \quad-3\rho^{2}F_{2}(\rho)-4\left(\left(\rho^{4}+1\right)F_{1}^{\prime}\right)^{\prime}
			+\frac{2\left(1-\rho^{8}\right)^{\nicefrac{3}{4}}}{\rho^{3}\left(1-\rho^{4}\right)^{\nicefrac{5}{8}}}
			\left(\rho\frac{\left(1-\rho^{8}\right)^{\nicefrac{7}{8}}}
			{\left(1+\rho^{4}\right)^{\nicefrac{5}{8}}}\,F_{1}^{\prime}{}^{2}\right)^{\prime}\Biggr]\\
	b^{-2}\mathbf{P}_{2}^{(2)}(\rho) &= 2\rho+4\rho^{3/2}\left(\rho^{3/2}F_{1}\right)^{\prime}
		+\theta\Biggl[-4\rho-\frac{8}{\rho^{3}}+4\rho^{3/2}\left(\rho^{3/2}\left(2F_{1}+F_{2}\right)\right)^{\prime}
		-2\left(\frac{8+3\rho^{4}}{\rho}\right)^{\nicefrac{1}{2}}\left(\left(\frac{8+3\rho^{4}}{\rho}\right)^{\nicefrac{1}{2}}F_{1}\right)^{\prime}\Biggr]\\
	b^{-2}\mathbf{P}_{3}^{(2)}(\rho) &= 4\rho+\frac{4}{\rho^{3}}+\theta\Biggl[8 \frac{33+\rho^{4}-\rho^{8}}{\rho^{7}}\Biggr]\,.
\end{flalign*}
\end{widetext}
The solution to the equation of motion \eqref{E:EOMs}  takes the form
\begin{equation}
	\Pi_{\mu\nu}^{(2)}=\sum_{j=0}^{3}\Lambda_{j}\Sigma_{\mu\nu}^{(j)}
\end{equation}
where
\begin{align}
	\Lambda_{j}=-\int_{\rho}^{\infty}\left\{ \frac{\int_{1}^{x}\mathbf{P}_{j}^{(2)}\left(x^{\prime}\right)dx^{\prime}}{x\left(x^{4}-1\right)}\left[1
		+2\theta\frac{2x^{4}-3}{x^{4}\left(x^{4}-1\right)}\right]\right\} dx\,.
\end{align}

To complete the calculation we need to insert the coefficient of the fourth order term in the series expansion of $\Pi_{\mu\nu}^{(2)}$ around $r=\infty$ into \eqref{E:Tmnprescription2}. Recalling that $8\theta = \delta = (c-a)/c$ we arrive at
\begin{align}
\begin{split}
	\lambda_0 &= \frac{\pi^2 T^2}{32 \pi G_5} \left[ 2-\log 2 + \frac{1}{8}\delta\left(-21 + \log 32\right)  \right] \\
	\lambda_1 &= \frac{\pi^2 T^2}{32 \pi G_5} \left[ 1 - \frac{7}{8} \delta  \right] \\
	\lambda_2 &= \frac{\pi^2 T^2}{32 \pi G_5} \left[ 2\log 2 + \frac{1}{8}\delta(14 - 2 \log 32) \right] \\
	\lambda_3 &= -\frac{\pi^2 T^2}{32 \pi G_5} 14 \delta
\end{split}
\end{align}
from which relation \eqref{E:linear} follows.
	
\vskip 1 mm
\textit{Acknowledgements}---We thank N.~Banerjee, S.~Cremonini, K.~Jensen and R.~Loganayagam for useful discussions. AY~is a Landau fellow, supported in part by the Taub foundation. ES and AY are supported in part by the ISF under grant number 495/11, by the BSF under grant number 2014350  and by the European commission FP7, under IRG 908049.

\bibliography{GB2nd}
	
\end{document}